# Aneka: A Software Platform for .NET-based Cloud Computing

Christian VECCHIOLA[a], Xingchen CHU[a,b], and Rajkumar BUYYA[a,b,1]
[a] *Grid Computing and Distributed Systems (GRIDS) Laboratory*
*Department of Computer Science and Software Engineering*
*The University of Melbourne, Australia*
[b] *Manjrasoft Pty Ltd, Melbourne, Australia*

**Abstract.** Aneka is a platform for deploying Clouds developing applications on top of it. It provides a runtime environment and a set of APIs that allow developers to build .NET applications that leverage their computation on either public or private clouds. One of the key features of Aneka is the ability of supporting multiple programming models that are ways of expressing the execution logic of applications by using specific abstractions. This is accomplished by creating a customizable and extensible service oriented runtime environment represented by a collection of software containers connected together. By leveraging on these architecture advanced services including resource reservation, persistence, storage management, security, and performance monitoring have been implemented. On top of this infrastructure different programming models can be plugged to provide support for different scenarios as demonstrated by the engineering, life science, and industry applications.

## Introduction

With the advancement of the modern human society, basic and essential services are delivered almost to everyone in a completely transparent manner. Utility services such as water, gas, and electricity have become fundamental for carrying out our daily life and are exploited on a pay per use basis. The existing infrastructures allow delivering such services almost anywhere and anytime so that we can simply switch on the light, open the tap, and use the stove. The usage of these utilities is then charged, according to different policies, to the end user. Recently, the same idea of utility has been applied to computing and a consistent shift towards this approach has been done with the spread of Cloud Computing.

Cloud Computing [1] is a recent technology trend whose aim is to deliver on demand IT resources on a pay per use basis. Previous trends were limited to a specific class of users, or focused on making available on demand a specific IT resource, mostly computing. Cloud Computing aims to be global and to provide such services to the masses, ranging from the end user that hosts its personal documents on the Internet, to enterprises outsourcing their entire IT infrastructure to external data centers. Never before an approach to make IT a real utility has been so global and complete: not only computing and storage resources are delivered on demand but the entire stack of computing can be leveraged on the Cloud.

[1] Corresponding Author.

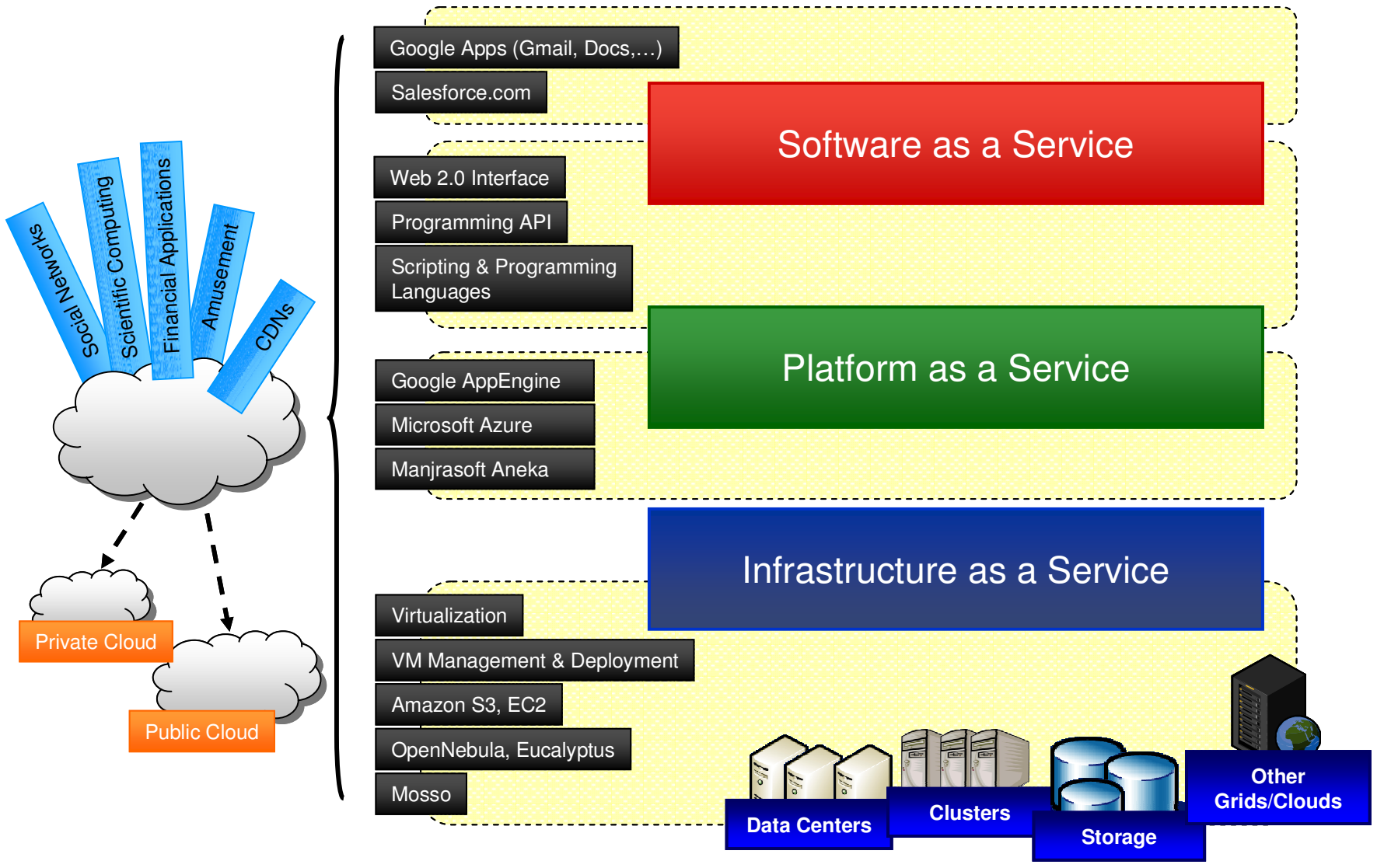

**Figure 1.** Cloud Computing architecture.

Figure 1 provides an overall view of the scenario envisioned by Cloud Computing. It encompasses so many aspects of computing that very hardly a single solution is able to provide everything that is needed. More likely, specific solutions can address the user needs and be successful in delivering IT resources as a real utility. Figure 1 also identifies the three pillars on top of which Cloud Computing solutions are delivered to end users. These are: *Software as a Service (SaaS)*, *Platform as a Service (PaaS)*, and *Infrastructure/Hardware as a Service (IaaS/HaaS).* These new concepts are also useful to classify the available options for leveraging on the Cloud the IT needs of everyone. Examples of Software as a Service are Salesforce.com[2] and Clarizen.com[3], which respectively provide on line CRM and project management services. PaaS solutions, such as Google AppEngine[4], Microsoft Azure[5], and Manjrasoft Aneka provide users with a development platform for creating distributed applications that can automatically scale on demand. Hardware and Infrastructure as a Service solutions provide users with physical or virtual resources that are fitting the requirements of the user applications in term of CPU, memory, operating system, and storage. These and any others QoS parameters are established through a Service Level Agreement (SLA) between the customer and the provider. Examples of this approach are Amazon EC2[6] and S3[7], and Mosso[8].

---

[2] http://www.salesforce.com
[3] http://www.clarenz.com
[4] http://code.google.com/appengine/docs/whatisgoogleappengine.html
[5] http://www.microsoft.com/azure/

It is very unlikely that a single solution provides the complete stack of software, platform, infrastructure and hardware as a service. More commonly, specific solutions provide services at one (or more) of these layers in order to exploit as many as possible the opportunities offered by Cloud Computing. Within this perspective, Aneka provides a platform for developing distributed applications that can easily scale and take advantage of Cloud based infrastructures. Aneka is software framework based on the .NET technology initially developed within the Gridbus project [2] and then commercialized by Manjrasoft[9]. It simplifies the development of distributed applications by providing: a collection of different ways for expressing the logic of distributed applications, a solid infrastructure that takes care of the distributed execution of applications, and a set of advanced features such as the ability to reserve and price computation nodes and to integrate with existing cloud infrastructures such as Amazon EC2.

This chapter provides an overview of Aneka as a framework for developing distributed applications and we will underline those features that make Aneka a Platform as a Service solution in the Cloud Computing model. The remainder of this chapter is organized as follows: Section 1 provides a brief introduction to the Cloud Computing architecture and features a comparison between some available commercial options. Section 2 gives an overview of Aneka by describing its service oriented architecture and the fundamental components of the system such as the Container and the core services. Section 3 presents application development with Aneka. In particular, the different Programming Models supported by the framework and the Software Development Kit are addressed. Section 4 provides an overview of the tools available within Aneka to manage the system, deploy applications, and monitor their execution. Section 5 describes some case studies where Aneka has been used to address the needs of scalability for different classes of applications. Conclusions and a discussion about the future development directions follow in Section 6.

## 1. Cloud Computing Reference Model and Technologies

In order to introduce a reference model for Cloud Computing, it is important to provide some insights on the definition of the term Cloud. There is no univocally accepted definition of the term. Fox et al. [3] notice that *"Cloud Computing refers to both the applications delivered as services over the Internet and the hardware and system software in the datacenters that provide those services"*. They then identify the Cloud with both the datacenter hardware and the software. A more structured definition is given by Buyya et al. [4] who define a Cloud as a *"type of parallel and distributed system consisting of a collection of interconnected and virtualized computers that are dynamically provisioned and presented as one or more unified computing resources based on service-level agreement"*. As it can be noticed, there is an agreement on the fact that Cloud Computing refers to the practice of delivering software and infrastructure as a service, eventually on a pay per use basis. In the following, we will illustrate how this is accomplished by defining a reference model for Cloud Computing.

[6] http://aws.amazon.com/ec2/
[7] http://aws.amazon.com/s3/
[8] http://www.mosso.com/
[9] http://www.manjrasoft.com/

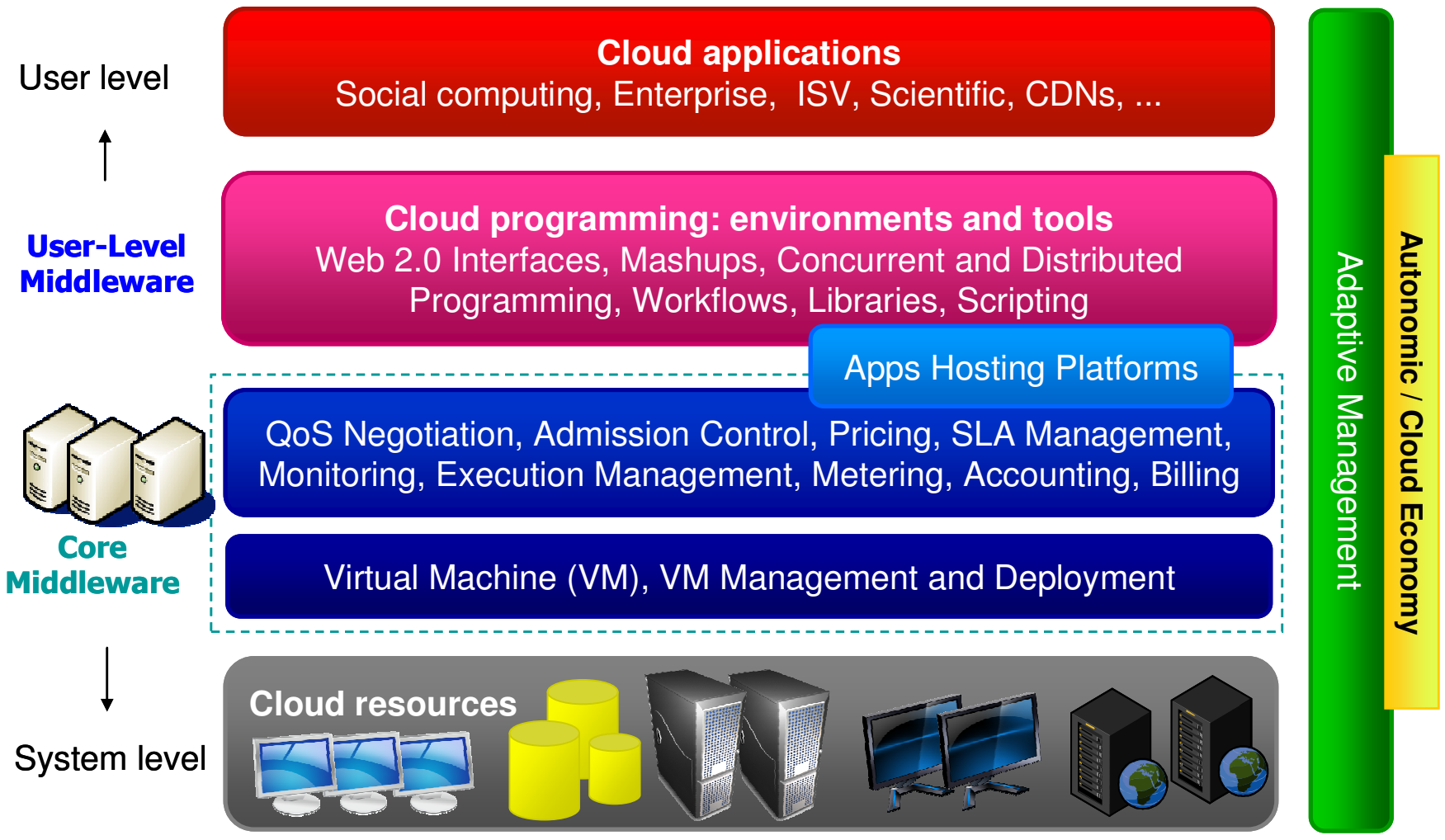

**Figure 2.** Cloud Computing layered architecture.

Figure 2 gives a layered view of the Cloud Computing stack. It is possible to distinguish four different layers that progressively shift the point of view from the system to the end user. The lowest level of the stack is characterized by the physical resources on top of which the infrastructure is deployed. These resources can be of different nature: clusters, datacenters, and spare desktop machines. Infrastructure supporting commercial Cloud deployments are more likely to be constituted by datacenters hosting hundreds or thousands of machines, while private Clouds can provide a more heterogeneous scenario in which even the idle CPU cycles of spare desktop machines are used to leverage the compute workload. This level provides the "horse power" of the Cloud.

The physical infrastructure is managed by the core middleware layer whose objectives are to provide an appropriate run time environment for applications and to exploit at best the physical resources. In order to provide advanced services, such as application isolation, quality of service, and sandboxing, the core middleware can rely on virtualization technologies. Among the different solutions for virtualization, hardware level virtualization and programming language level virtualization are the most popular. Hardware level virtualization guarantees complete isolation of applications and a fine partitioning of the physical resources, such as memory and CPU, by means of virtual machines. Programming level virtualization provides sandboxing and managed execution for applications developed with a specific technology or programming language (i.e. Java, .NET, and Python). On top of this, the core middleware provides a wide set of services that assist service providers in delivering a professional and commercial service to end users. These services include:

negotiation of the quality of service, admission control, execution management and monitoring, accounting, and billing.

Together with the physical infrastructure the core middleware represents the platform on top of which the applications are deployed in the Cloud. It is very rare to have direct access to this layer. More commonly, the services delivered by the core middleware are accessed through a user level middleware. This provides environments and tools simplifying the development and the deployment of applications in the Cloud: web 2.0 interfaces, command line tools, libraries, and programming languages. The user level middleware constitutes the access point of applications to the Cloud.

The Cloud Computing model introduces several benefits for applications and enterprises. The adaptive management of the Cloud allows applications to scale on demand according to their needs: applications can dynamically acquire more resource to host their services in order to handle peak workloads and release when the load decreases. Enterprises do not have to plan for the peak capacity anymore, but they can provision as many resources as they need, for the time they need, and when they need. Moreover, by moving their IT infrastructure into the Cloud, enterprise can reduce their administration and maintenance costs. This opportunity becomes even more appealing for startups, which can start their business with a small capital and increase their IT infrastructure as their business grows. This model is also convenient for service providers that can maximize the revenue from their physical infrastructure. Besides the most common "pay as you go" strategy more effective pricing policies can be devised according to the specific services delivered to the end user. The use of virtualization technologies allows a fine control over the resources and the services that are made available at runtime for applications. This introduces the opportunity of adopting various pricing models that can benefit either the customers or the vendors.

The model endorsed by Cloud Computing provides the capability of leveraging the execution of applications on a distributed infrastructure that, in case of public clouds, belongs to third parties. While this model is certainly convenient, it also brings additional issues from a legal and a security point of view. For example, the infrastructure constituting the Computing Cloud can be made of datacenters and clusters located in different countries where different laws for digital content apply. The same application can then be considered legal or illegal according to the where is hosted. In addition, privacy and confidentiality of data depends on the location of its storage. For example, confidentiality of accounts in a bank located in Switzerland may not be guaranteed by the use of data center located in United States. In order to address this issue some Cloud Computing vendors have included the geographic location of the hosting as a parameter of the service level agreement made with the customer. For example, Amazon EC2 provides the concept of availability zones that identify the location of the datacenters where applications are hosted. Users can have access to different availability zones and decide where to host their applications. Since Cloud Computing is still in its infancy the solutions devised to address these issues are still being explored and will definitely become fundamental when a wider adoption of this technology takes place.

Table 1 identifies some of the major players in the field and the kind of service they offer. Amazon Elastic Compute Cloud (EC2) operates at the lower levels of the Cloud Computing reference model. It provides a large computing infrastructure and a service based on hardware virtualization. By using the Amazon Web Services users can create Amazon Machine Images (AMIs) and save them as templates from which multiple instances can be run. It is possible to run either Windows or Linux virtual machines and the user is charged per hour for each of the instances running. Amazon also provides storage services with the Amazon Simple Storage Service (S3), users can take advantage of Amazon S3 to move large data files into the infrastructure and get access to them from virtual machine instances.

**Table 1.** Feature comparison of some of the commercial offerings for Cloud Computing.

| Properties | Amazon EC2 | Google AppEngine | Microsoft Azure | Manjrasoft Aneka |
|---|---|---|---|---|
| Service Type | IaaS | IaaS – PaaS | IaaS – PaaS | PaaS |
| Support for (value offer) | Compute/storage | Compute (web applications) | Compute | Compute |
| Value added service provider | Yes | Yes | Yes | Yes |
| User access interface | Web APIs and Command Line Tools | Web APIs and Command Line Tools | Azure Web Portal | Web APIs, Custom GUI |
| Virtualization | OS on Xen hypervisor | Application Container | Service Container | Service Container |
| Platform (OS & runtime) | Linux, Windows | Linux | .NET on Windows | .NET/Mono on Windows, Linux, MacOS X |
| Deployment model | Customizable VM | Web apps (Python, Java, JRuby) | Azure Services | Applications (C#, C++, VB, ….) |
| If PaaS, ability to deploy on 3rd pathy IaaS | N.A. | No | No | Yes |

While the commercial offer of Amazon can be classified completely as a IaaS solutions, Google AppEngine, and Microsoft Azure are integrated solutions providing both a computing infrastructure and a platform for developing applications. Google AppEngine is a platform for developing scalable web applications that will be run on top of server infrastructure of Google. It provides a set of APIs and an application model that allow developers to take advantage of additional services provided by Google such as Mail, Datastore, Memcache, and others. By following the provided application model, developers can create applications in Java, Python, and JRuby. These applications will be run within a sandbox and AppEngine will take care of automatically scaling when needed. Google provides a free limited service and utilizes daily and per minute quotas to meter and price applications that require a professional service.

Azure is the solution provided by Microsoft for developing scalable applications for the Cloud. It is a cloud services operating system that serves as the development, run-time, and control environment for the Azure Services Platform. By using the Microsoft Azure SDK developers can create services that leverage on the .NET Framework. These services are then uploaded to the Microsoft Azure portal and executed on top of Windows Azure. Microsoft Azure provides additional services such as workflow execution and management, web services orchestration, and access to SQL data stores. Currently, Azure is still in Community Technical Preview and its usage is free, its commercial launch is scheduled for the second half of 2009 and users will be charged by taking into account the CPU time, the bandwidth and the storage used, the number of transaction performed by their services, and also the use of specific services such as SQL or .NET services.

Differently from all the previous solutions, Aneka is a pure implementation of the Platform as a Service model. The core value of Aneka is a service oriented runtime environment that is deployed on both physical and virtual infrastructures and allows the execution of applications developed with different application models. Aneka provides a Software Development Kit (SDK) allowing developers to create cloud applications on any language supported by the .NET runtime and a set of tools for quickly setting up and deploying clouds on Windows and Linux based systems. Aneka can be freely downloaded and tried for a limited period, while specific arrangements have to be made with Manjrasoft for commercial use. In the remainder of this chapter we illustrate the features of Aneka.

## 2. Aneka Architecture

Aneka is a platform and a framework for developing distributed applications on the Cloud. It harnesses the spare CPU cycles of a heterogeneous network of desktop PCs and servers or datacenters on demand. Aneka provides developers with a rich set of APIs for transparently exploiting such resources and expressing the business logic of applications by using the preferred programming abstractions. System administrators can leverage on a collection of tools to monitor and control the deployed infrastructure. This can be a public cloud available to anyone through the Internet, or a private cloud constituted by a set of nodes with restricted access.

Aneka is based on the .NET framework and this is what makes it unique from a technology point of view as opposed to the widely available Java based solutions. While mostly designed to exploit the computing power of Windows based machines, which are most common within an enterprise environment, Aneka is portable over different platforms and operating systems by leveraging other implementations of the ECMA 334 [5] and ECMA 335 [6] specifications such as Mono. This makes Aneka an interesting solution for different types of applications in educational, academic, and commercial environments.

### *2.1. Overview*

Figure 3 gives an overview of the features of Aneka. The Aneka based computing cloud is a collection of physical and virtualized resources connected through a network, which could be the Internet or a private intranet. Each of these resources hosts an instance of the Aneka Container representing the runtime environment in which the distributed applications are executed. The container provides the basic management features of the single node and leverages all the other operations on the services that it is hosting. In particular we can identify fabric, foundation, and execution services. Fabric services directly interact with the node through the Platform Abstraction Layer (PAL) and perform hardware profiling and dynamic resource provisioning. Foundation services identify the core system of the Aneka middleware, they provide a set of basic features on top of which each of the Aneka containers can be specialized to perform a specific set of tasks. Execution services directly deal with the scheduling and execution of applications in the Cloud. One of the key features of Aneka is the ability of providing different ways for expressing distributed applications by offering different programming models; execution services are mostly concerned with providing the middleware with an implementation for these models. Additional services such as persistence and security are transversal to the entire stack of services that are hosted by the Container. At the application level, a set of different components and tools are provided to: 1) simplify the development of applications (SDK); 2) porting existing applications to the Cloud; and 3) monitoring and managing the Aneka Cloud.

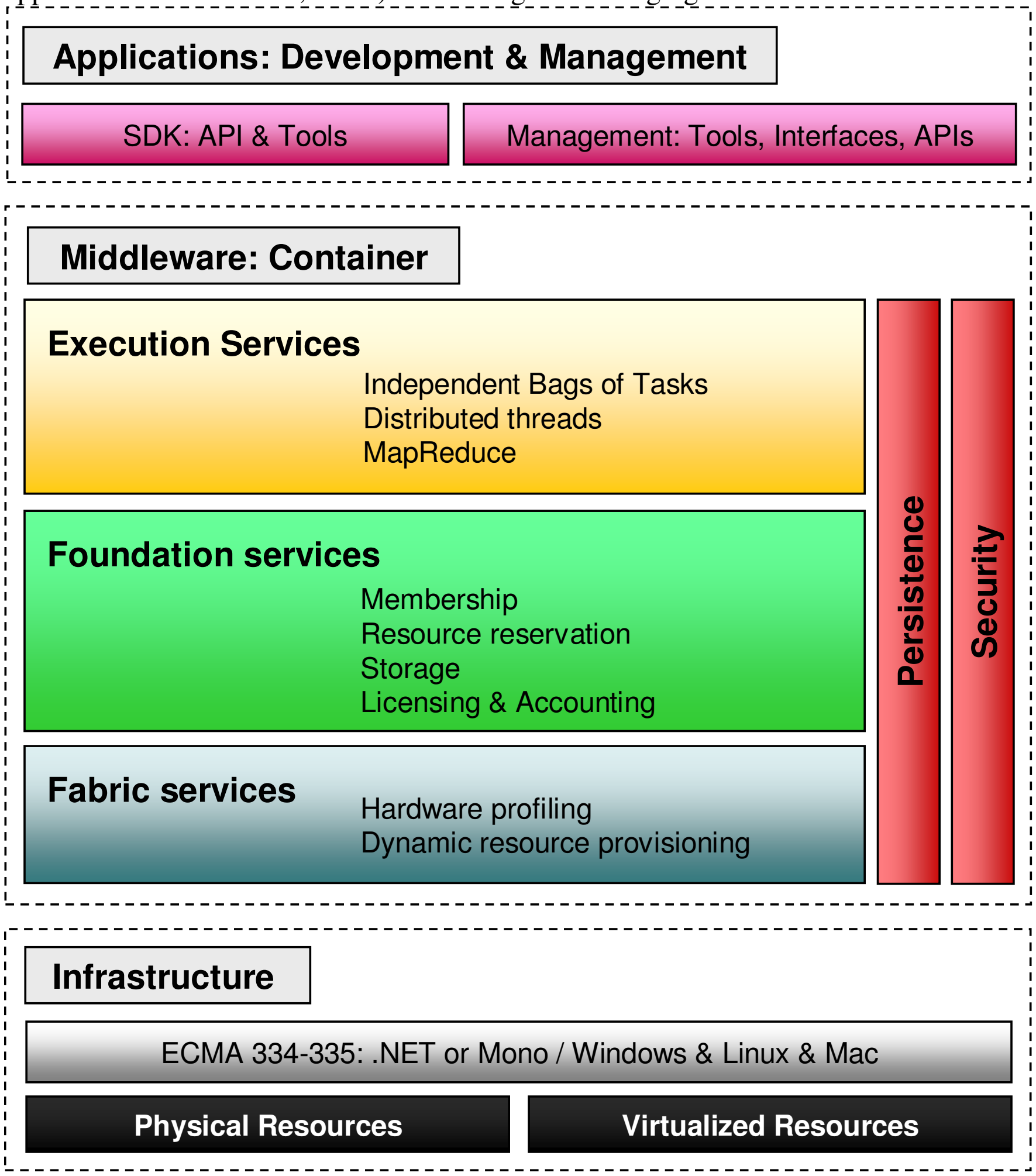

**Figure 3.** Overview of the Aneka framework.

A common deployment of Aneka is presented in Figure 4. An Aneka based Cloud is constituted by a set of interconnected resources that are dynamically modified according to the user needs by using resource virtualization or by harnessing the spare CPU cycles of desktop machines. If the deployment identifies a private Cloud all the resources are in house, for example within the enterprise. This deployment is extended by adding publicly available resources on demand or by interacting with other Aneka public clouds providing computing resources connected over the Internet. The heart of this infrastructure is the Aneka Container which represents the basic deployment unit of Aneka based clouds.

Some of the most characteristic features of the Cloud Computing model are:

- flexibility,
- elasticity (scaling up or down on demand), and
- pay per usage.

The architecture and the implementation of the Container play a key role in supporting these three features: the Aneka cloud is flexible because the collection of services available on the container can be customized and deployed according to the specific needs of the application. It is also elastic because it is possible to increase on demand the number of nodes that are part of the Aneka Cloud according to the user needs. The integration of virtual resources into the Aneka Cloud does not introduce specific challenges: once the virtual resource is acquired by Aneka it is only necessary to have an administrative account and a network access to it and deploy the Container on it as it happens for any other physical node. Moreover, because of the Container being the interface to hosting node it is easy to monitor, meter, and charge any distributed application that runs on the Aneka Cloud.

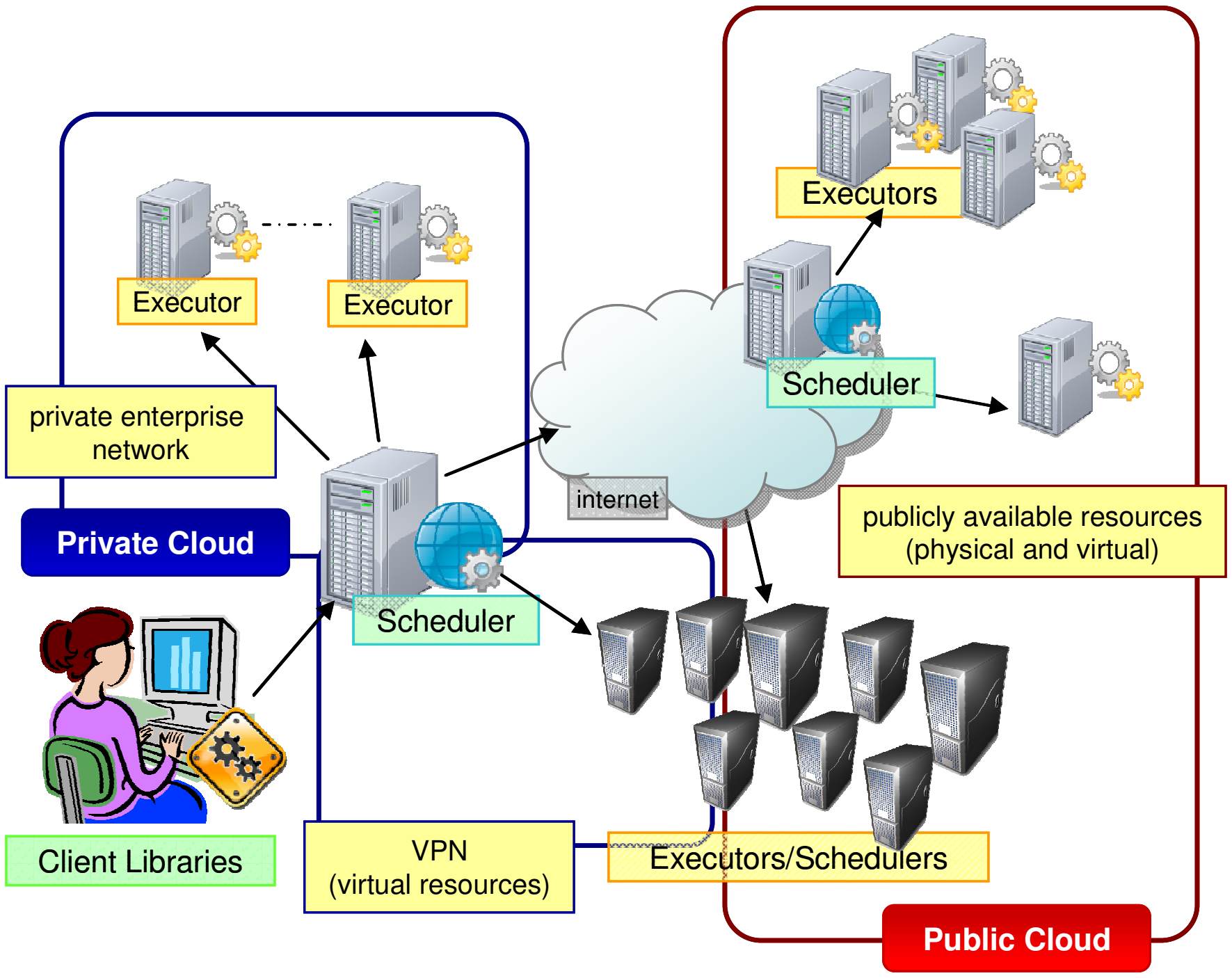

**Figure 4.** Deployment scenario for Aneka

### *2.2. Anatomy of the Aneka Container*

The Container represents the basic deployment unit of Aneka based Clouds. The network of containers defining the middleware of Aneka constitutes the runtime environment hosting the execution of distributed applications. Aneka strongly relies on a Service Oriented Architecture [7] and the Container is a lightweight component providing basic node management features. All the other operations that are required by the middleware are implemented as services.

Figure 3 illustrates the stack of services that can be found in a common deployment of the Container. It is possible to identify four major groups of services:

- Fabric Services
- Foundation Services
- Execution Services
- Transversal Services

The collective execution of all these services actually creates the required runtime environment for executing applications. Fabric services directly interface with the hosting resource and are responsible for low level operations, foundation services constitute the core of the runtime environment, and execution services manage the execution of applications. A specific class – Transversal Services – operates at all levels and provides support for security and persistence.

Additional and specific services can be seamlessly integrated into the Container by simply updating a configuration file. This operation can be performed either by means of an automated procedure or manually. The ability of hosting on demand new services and unloading existing services makes the Aneka Container an extremely configurable component able to address and elastically react to the changing needs of the applications by scaling up or down the set of services installed in the system. Moreover, by relying on services and message passing for implementing all the features of the system, the Aneka Container can easily evolve and integrate new features with minimum setup costs.

### *2.3. Fabric Services*

Fabric services define the lowest level of the software stack representing the Aneka Container. They provide access to the resource provisioning subsystem and to the hardware of the hosting machine. Resource provisioning services are in charge of dynamically providing new nodes on demand by relying on virtualization technologies, while hardware profile services provide a platform independent interface for collecting performance information and querying the properties of the host operating system and hardware.

Hardware profiling services provide a platform independent interface for accessing the operating system and the underlying hardware. These services rely on the Platform Abstraction Layer (PAL) that allows the Container to be completely independent from the hosting machine and the operating system and the whole framework to be portable over different platforms. In particular the following information is collected for all the supported runtimes and platforms:

- Static and dynamic CPU information (CPUs, operating frequency, CPU usage);
- Static and dynamic memory information (size, available, and used);
- Static and dynamic storage information (size, available, and used);

This information is collected for each of the nodes belonging to the Aneka Cloud and made available to the other services installed in the systems. For example, execution services and in particular scheduling components, can take advantage of dynamic performance information to devise a more efficient scheduling for applications.

Dynamic resource provisioning allows the Aneka Cloud to elastically scale up and down according to the requirements of applications. These services are in charge of dynamically acquiring and integrating new nodes into the Aneka Cloud in order to satisfy the computation needs of one or more applications. Dynamic resource provisioning addresses two different scenarios: physical resource provisioning and virtual resource provisioning. With physical resource provisioning one Aneka Cloud simply "borrows" some nodes from other Aneka Clouds by specifying a service level agreement and the specific characteristics required for these nodes in terms of services and hardware. With virtual resource provisioning the nodes are dynamically acquired by interacting with existing virtual machine managers or IaaS implementations such as Amazon EC2 or Amazon S3. In this case, the Aneka Cloud requests as many virtual machines as needed to deploy an Aneka Container together with the required services. The way in which new resources are integrated into the Cloud characterizes the type of Cloud managed by Aneka. If resources are collected from a private internal network either via a hypervisor or another Aneka Cloud, the resulting system is still a private Cloud. If resources are obtained by relying on a publicly available Aneka Cloud, the entire system may be a public or hybrid Cloud. We have a public Cloud if the initial system was a public Cloud, a hybrid Cloud otherwise.

Resource provisioning and hardware profiling are fundamental in a Cloud environment where resources are obtained on demand and subject to specific service level agreements. In particular resource reservation strongly relies on the information obtained by these services. Aneka allows reserving nodes for a specific application. It is possible to specify the set of characteristics required for each of these nodes, and the number of nodes. The reservation service will then, if possible, reserve within the Aneka Cloud those nodes that fulfill the requirements requested by the application. To accomplish this it is necessary to access to the static and dynamic performance information of the node. Advanced strategies can then rely on dynamic resource provisioning in order to make up for the lack of resources.

### *2.4. Foundation Services*

Together with the fabric services the foundation services represent the core of the Aneka middleware on top of which Container customization takes place. Foundation services constitute the pillars of the Aneka middleware and are mostly concerned with providing runtime support for execution services and applications. The core of Aneka addresses different issues:

- Directory and Membership;
- Resource reservation;

- Storage management;
- Licensing, accounting, and pricing;

These services can be directly consumed by users, applications, or execution services. For example, users or applications can reserve nodes for execution, while execution services can query the Membership Catalogue in order to discover whether the required services are available in the Cloud to support the execution of a specific application. Licensing, accounting, and pricing are services that will be more of interest for single users or administrators.

### *2.4.1. Directory and Membership*

Directory and Membership Services are responsible for setting up and maintaining the information about the nodes and the services constituting the Aneka Cloud. These services include *Membership Catalogue*, *Heartbeat Service*, and *Discovery Service*. The Membership Catalogue acts as global directory maintaining the list of available services and their location in the Aneka Cloud. The information in the Membership Catalogue is dynamically updated by the Heartbeat Services installed in each node belonging to the Cloud. The Heartbeat services collect the statistic information about the hosting node from the Hardware profiling services and update the Membership Catalogue periodically. The Aneka middleware exposes some autonomic properties [8] being able not only to react to failures but also to auto-configure itself when connections between nodes are broken and nodes are not reachable. This ability is mostly provided by the Discovery Service, which is in charge of discovering the available Aneka nodes on the Cloud and providing the required information for adding a node to the Membership Catalogue. The collective execution of these three services allows the automatic setting up of an Aneka Cloud without any static configuration information, but simply an available network connection.

### *2.4.2. Resource Reservation*

Resource reservation is a fundamental feature in any distributed middleware aiming to support application execution with a specific quality of service (QoS). Resource reservation identifies the ability of reserving a set of nodes and using them for executing a specific application. Without such capability, it is impossible to guarantee many of the most important QoS parameters, since it is not possible to control the execution of applications. Aneka provides an advanced reservation infrastructure that works across almost all the supported programming models, that allows users to reserve a collection of nodes for a given time frame, and assign this reservation to a specific application. The infrastructure guarantees that at the time specified within the reservation the selected resources are made available for executing the application.

In order to support the ability of reserving compute resources two different components have been implemented: *Reservation Service* and *Allocation Manager*. The Reservation Service is a central service that keeps track of the allocation map of all the nodes constituting the Aneka Cloud, while the Allocation Manager provides a view of the allocation map of the local Container. The Reservation Service and the Allocation Manager Services deployed in every Container provide the infrastructure that enables to reservation of compute resources, and guarantee the desired QoS. During application execution a collection of jobs are submitted to the Aneka Cloud and each of these jobs are actually moved and executed in the runtime environment set up by the Container on a specific resource. Reserved nodes only accept jobs that belong to the reservation request that is currently active. In case there is no active reservation on the node any job that matches the security requirements set by Aneka Cloud is executed. The Allocation Manager is responsible for keeping track of the reserved time frames in the local node and of checking – before the execution of jobs start – whether they are admissible or not. The Reservation Service is indeed responsible for providing a global view to the execution services and users of the status of the system, and, by interacting with the cloud schedulers, for implementing a reservation aware application execution.

In a cloud environment, the ability of reserving resources for application execution is fundamental, not only because it offers a ways for guaranteeing the desired QoS, but also because it provides an infrastructure to implement pricing mechanisms. Aneka provides some advanced features integrated within the Reservation Service that allow a flexible pricing scheme for applications. In particular it implements the *alternate offers protocol* [9], which allows the infrastructure to provide the user with a counter offer in case the QoS parameters of the initial request cannot be met by the system. This feature, together with the ability of dynamically provisioning additional nodes for computation, makes the reservation infrastructure a key and innovative characteristic of Aneka.

#### *2.4.3. Storage management*

The availability of disk space, or more generally storage, is a fundamental feature for any distributed system implementation. Applications normally require files to perform their tasks, whether they are data files, configuration files, or simply executable files. In a distributed context these files have to be moved – or at least made reachable from – where the execution takes place. These tasks are normally carried out by the infrastructure representing the execution middleware and in a cloud environment these operations become even more challenging because of the dynamic nature of the system.

In order to address this issue Aneka implements a Storage Service. This service is responsible for providing persistent, robust, file based storage for applications. It constitutes a staging facility for all the nodes belonging to the Aneka Cloud and also performs data transfers among Aneka nodes, the client machine, and remote servers. In a cloud environment the user requirements can be different and dynamically change during the lifetime of the applications. Such requirements can also affect storage management in terms of their location and of the specific media used to transfer information. Aneka provides an infrastructure able to support a different set of storage facilities. The current release of Aneka provides a storage implementation based on the File Transfer Protocol (FTP) service. Additional storage facilities can be integrated into the system by providing a specific implementation of a *data channel*. A data channel represents the interface used within Aneka to access a specific storage facility. Each data channel implementation consists of a server component, that manages the storage space made available with the channel, and a client component, which is used to remotely access that space. Aneka can transparently plug any storage facility for which a data channel implementation has been provided and transparently use it. The use of data channels is transparent to users too, who simply specify the location of the files needed by their application and the protocol through which they are made accessible. Aneka will automatically the system with the components needed to import the required files into the Cloud.

The architecture devised to address storage needs in Aneka provides a great flexibility and extensibility. Not only different storage facilities can be integrated but they also can be composed together in order to move data across different mediums and protocols. This allows Aneka Clouds a great level of interoperability from the perspective of data.

### *2.4.4. Licensing, Accounting, and Pricing*

Aneka provides an infrastructure that allows setting up public and private clouds. In a cloud environment, especially in the case of public clouds, it is important to implement mechanisms for controlling resources and pricing their usage in order to charge users. Licensing, accounting, and pricing are the tasks that collectively implement a pricing mechanism for applications in Aneka.

The *Licensing Service* provides the very basic resource controlling feature that protects the system from misuse. It restricts the number of resources that can be used for a certain deployment. Every container that wants to join the Aneka Cloud is subject to verification against the license installed in the system and its membership is rejected if restrictions apply. These restrictions can involve the number of maximum nodes allowed in the Aneka Cloud, or a specific set of services hosted by the container. This service does not provide any direct benefit for users but prevent the system from malicious system administrators that want to overprovision the Aneka Cloud.

The *Accounting* and *Pricing Services*, available in the next release of Aneka, are more directly related with billing the user for using the Cloud. In particular the Accounting Service keeps track of applications running, their reservations, and of the users they belong to, while the Pricing Service is in charge of providing flexible pricing strategies that benefit both the users of the Cloud and the service providers. These two components become important in case of dynamic resource provisioning of virtual resources: IaaS implementations such as Amazon EC2 charge the usage of the virtual machines per hour. The way in which the cost of this service is reflected into the user bill is the responsibility of the Pricing Service.

### *2.5. Execution Services*

Execution services identify the set of services that are directly involved in the execution of distributed applications in the Aneka Cloud. The application model enforced by Aneka represents a distributed application as a collection of jobs. For any specific programming model implemented in Aneka at least two components are required providing execution support: *Scheduling Service* and *Execution Service*. The Scheduling Service coordinates the execution of applications in the Aneka Cloud and is responsible for dispatching the collection of jobs generated by applications to the compute nodes. The Execution Service constitutes the runtime environment in which jobs are executed. More precisely, it is in charge of retrieving all the files required for execution, monitoring the execution of the job, and collecting the results. The number and the type of services required to deploy a programming model varies according to the specific nature of the programming model. Generally these two services are the only ones required in most of the cases. The Task Model, the Thread Model, and the MapReduce Model are implemented according to this scheme.

Execution Services can then rely on other existing services, available with a common deployment of the Aneka Cloud, to provide a better support for application execution. For example they can integrate with the existing Reservation Service and Storage service to support quality of service for application execution and support for data transfer. The integration with these services is completely dynamic and no static binding is required.

A common deployment scenario of an Aneka Cloud concentrates the scheduling services of all the programming models in one or few nodes, while configuring all the other nodes with execution services, thus creating a master-slave topology. Whereas this deployment is quite common, the service oriented architecture of Aneka does not enforce it and more balanced and dynamic topologies can be devised by system administrators. For example, environments characterized by thousands of machines can more easily scale and reconfigure by means of hierarchical topologies and brokering infrastructures. Hierarchical topologies can help in distributing the overload of managing huge number of resources: in this setting, the scheduling service managing a network of nodes where execution services are deployed. These scheduling services can be then seen as multi-core from other meta schedulers which coordinate the load of the system at a higher level. Such structure can be enhanced and made more dynamic by integrating into the Aneka Container brokering services that, by means of dynamic SLAs extend and enrich the set of features that are offered to the users of the Cloud. Other solutions [10], based on a peer to peer model, can also be implemented.

### *2.6. Transversal Services*

Aneka provides additional services that affect all the layers of the software stack implemented in the Container. For this reason they are called *transversal services*, such as the persistence layer and the security infrastructure.

#### *2.6.1. Persistence*

The persistence layer provides a complete solution for recording the status of the Cloud and for restoring it after a system crash or a partial failure. The persistence layer keeps track of the sensitive information for the distributed system such as: all the applications running in the Cloud and their status; the topology information of the Cloud and the current execution status; the status of the storage. This information is constantly updated and saved to the persistence storage. The persistence layer is constituted by a collection of persistence stores that are separately configured in order to provide the best required quality of service.

The current release of Aneka provides two different implementations for these components that can be used to configure and tune the performance of the Cloud:

- *In memory persistence*: this persistence model provides a volatile store that is fast and performing but not reliable. In case of system crash or partial failure the execution of the applications can be irreversibly compromised. While this solution is optimal for a quick setup of the Aneka Cloud and for testing purposes, it is not suggested for production deployments.
- *Relational Database*: this solution relies on the ADO.NET framework for providing a persistent store, which is generally represented by a database management system. In this case the information of the state of the Cloud and its components are saved inside database tables and retrieved when necessary. This solution provides reliability against failures and prevents from the loss of data but requires an existing installation of the supported RDBMS. The current implementation of Aneka supports two different backend for this kind of solution: MySQL 5.1 and SQL Server 2005 v9.0 onward.

These are just two ready to use implementations of the persistence layer. Third parties can provide a specific implementation and seamlessly integrate it into the systems with minimum effort. The possibilities for extending the system are many: it is possible to implement from scratch a new persistence layer or simply provide the SQL scripts that create tables and stored procedures for the database persistence layer.

### *2.6.2. Security*

The security layer provides access to the security infrastructure of Aneka. This layer separates authentication – that means identifying who users are – from authorization – that means what users are allowed to do. The implementation of these two functions relies on providers, which abstract the two operations within the framework, and user credentials, which contain the information required by the providers to authenticate and authorize users. Before any operation on behalf of the user is performed on the Aneka Cloud its credentials are verified against the authentication and authorization providers, which can reject or authorize the operation.

Specific implementations of these two providers can be seamlessly integrated into the infrastructure simply by editing the configuration of the Container. In this way it is possible to run Aneka on different security infrastructure according to specific requirements of the Cloud. Specific deployments can require the use of existing security infrastructures. In this case, the specific implementation of security providers will rely on the existing security model and user credentials will contain the required information for representing the user within the underlying security system. This has been the approach for supporting the Window Authentication in Aneka. In the case of Windows based deployments Aneka can rely on the Windows integrated security and provide access to the system for the specific Windows users. Alternatively, it is possible to set up a Cloud with no security at all, simply by using the Anonymous security providers, which do not perform any security check for user applications. Third parties can set up their own security providers by implementing the interfaces defined in the Aneka security APIs.

## *2.7. Portability and Interoperability*

Aneka is a Platform as a Service implementation of the Cloud Computing model and necessarily relies on the existing virtual and physical infrastructure for providing its services. More specifically, being developed on top of the Common Language Infrastructure, it requires an implementation of the ECMA 335 specification such as the .NET framework or Mono.

Since the Cloud is a dynamic environment aggregating heterogeneous computing resources, the choice of developing the entire framework on top of a virtual runtime environment, provides some interesting advantages. For example it is possible to easily support multiple platform and operating systems with reduced or no conversion costs at all. Developing for a virtual execution environment such as Java or the Common Language Infrastructure, does not necessarily mean to devise software that will naturally run on any supported platform. In the case of Aneka this aspect becomes even more challenging since some of the components of the framework directly interact with the hardware of the machine (physical or virtual) hosting the Aneka Container.

In order to address this issue a specific layer that encapsulates all the platform dependencies on the hosting platform behind a common interface has been integrated into Aneka. This layer is called Platform Abstraction Layer (PAL) and provides a unified interface for accessing all the specific properties of the Operating System and the underlying hardware that are of interest for Aneka. The PAL is a fundamental component of the system and constitutes the lowest layer of the software stack implemented in the Aneka Container. It exposes the following features:

- Uniform and platform independent interface for profiling the hosting platform;
- Uniform access to extended and additional properties of the hosting platform;
- Uniform and platform independent access to remote nodes;
- Uniform and platform independent management interfaces;

The dynamic and heterogeneous nature of computing clouds necessarily requires a high degree of flexibility in aggregating new resources. In the case of Aneka, adding one resource to the Cloud implies obtaining access to a physical or a virtual machine and deploying into it an instance of the Aneka Container. These operations are performed by the PAL, which not only abstracts the process for installing and deploying a new Container but also automatically configures it according to the hosting platform. At startup the container probes the system, detects the required implementation of the PAL, and loads it in memory. The configuration of the Container is performed in a completely transparent manner and makes its deployment on virtual and physical machines really straightforward.

The current release of Aneka provides a complete implementation of the PAL for Windows based systems on top of the .NET framework and for the Linux platform on top of Mono. A partial but working implementation of the PAL for Mac OS X based systems on top of Mono is also provided.

## 3. Application Development

Aneka is a platform for developing applications that leverage Clouds for their execution. It then provides a runtime infrastructure for creating public and private Clouds and a set of abstractions and APIs through which developers can design and implement their applications. More specifically Aneka provides developers with a set of APIs for representing the Cloud application and controlling their execution, and a set of *Programming Models* that are used to define the logic of the distributed application itself. These components are part of the Aneka Software Development Kit.

### *3.1. The Aneka SDK*

The Aneka Software Development Kit contains the base class libraries that allow developers to program applications for Aneka Clouds. Beside a collection of tutorials that thoroughly explain how to develop applications, the SDK contains a collection of class libraries constituting the *Aneka Application Model*, and their specific implementations for the supported programming models.

The Aneka Application Model defines the properties and the requirements for distributed applications that are hosted in Aneka Clouds. Differently from other middleware implementations Aneka does not support single task execution, but any unit of user code is executed within the context of a distributed application. An application in Aneka is constituted by a collection of execution units whose nature depends on the specific programming model used. An application is the unit of deployment in Aneka and configuration and security operates at application level. Execution units constitute the logic of the applications. The way in which units are scheduled and executed is specific to the programming model they belong to. By using this generic model, the framework provides a set of services that work across all programming model supported: storage, persistence, file management, monitoring, accounting, and security.

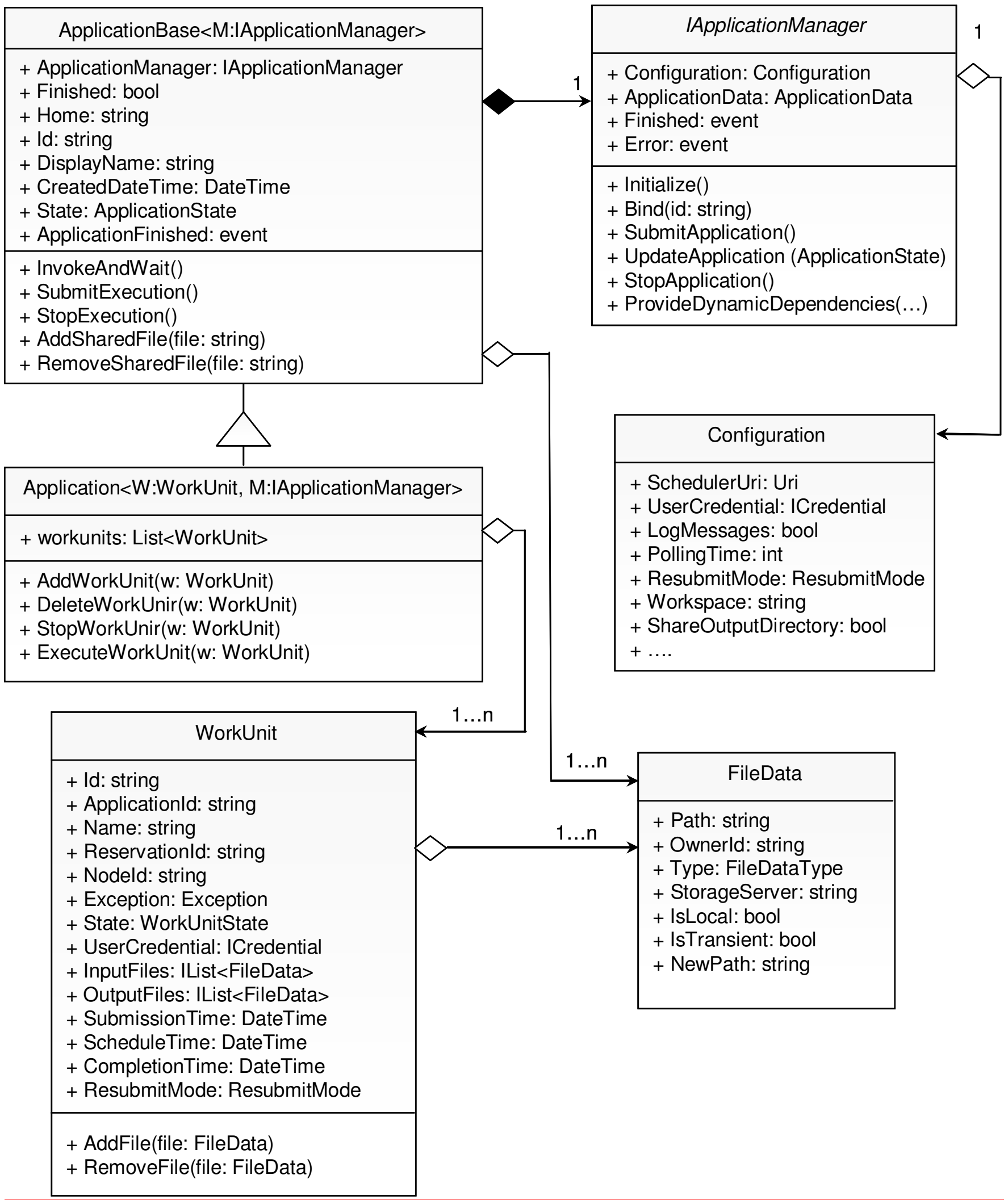

**Figure 5.** Aneka application model.

Figure 5 illustrates the key elements of the Aneka Application Model. As previously introduced an application is a collection of work units that are executed by the middleware. While the Application class contains the common operations for all the supported programming models, its template specialization customizes its behavior for a specific model. In particular each of the programming model implementations has to specify two types: the specific type of work unit and the specific type of application manager. The work unit represents the basic unit of execution of the distributed application, while the application manager is an internal component that is used to submit the work units to the middleware. The SDK provides base class implementations for these two types and developers can easily extend them and taking advantage of the services built for them.

The Software Development Kit also provides facilities for implementing the components required by the middleware for executing a programming model. In particular, it provides some base classes that can be inherited and extended for implementing schedulers and executors components. Developers that are interested in developing a new programming model can take as a reference the existing programming models and implement new models as a variation of them or they can completely from scratch by using the base classes. Moreover, the Aneka SDK also exposes APIs for implementing custom services that can be seamlessly plugged into the Aneka Container by editing its configuration file.

## *3.2. Programming Models*

A programming model represents a way for expressing a distributed application within Aneka. It defines the abstractions used by the user to model their application and the execution logic of these applications as a whole in the Aneka Cloud. Every application that is executed in the Aneka Cloud is expressed in terms of a specific programming model. The current release of Aneka includes three different programming models ready to use for developing applications. These are: *Task Programming Model*, *Thread Programming Model*, and *MapReduce Programming Model*.

### *3.2.1. Task Programming Model*

The *Task Programming Model* provides developers with the ability of expressing bag of tasks applications. By using the Task Model the user can create a distributed application and submit a collection of tasks to Aneka. The submission can be either static or dynamic. The scheduling and execution services will manage the execution of these tasks according to the available resources in the Aneka network.

Developers can use predefined tasks that cover the basic functionalities available from the OS shell or define new tasks by programming their logic. With tasks being independent from each other, this programming model does not enforce any execution order or sequencing but these operations have to be completely managed by the developer on the client application if needed.

The task programming model is the most straightforward programming model available with Aneka and can be used as a base on top of which other models can be implemented. For example the parameter sweeping APIs used by the Design Explorer rely on the Task Model APIs to create and submit the tasks that are generated for each of the combinations of parameters that need to be explored. More complex models such as workflows can take advantage of this simple and thin implementation for distributing the execution of tasks.

### *3.2.2. Thread Programming Model*

The *Thread Programming Model* allows quickly porting multi-threaded applications into a distributed environment. Developers familiar with threading API exposed by the .NET framework or Java can easily take advantage of the set of compute resources available with Aneka in order to improve the performance of their applications.

The Thread Model provides as fundamental component for building distributed applications the concept of distributed thread. A distributed thread exposes the same APIs of a thread in the .NET framework but is executed remotely. Developers familiar with the multi-threaded applications can create, start, join, and stop threads in the same way in which these operations are performed on local threads. Aneka will take care of distributing and coordinating the execution of these threads.

Compared to the Task Model the Thread Model provides a more complex, powerful, and lower level API. While the common usage for the Task Model is "submit and forget" – that means that users submit tasks and forget of their existence until they terminate – in the case of the Thread Model the developer is supposed to have a finer control on the single threads. This model is definitely the best option when a pre-existing multi-threaded application needs to be ported to a distributed environment for improving its performance. In this case minimal changes to the existing code have to be made to run such application by using the Thread Model.

#### *3.2.3. MapReduce Programming Model*

The *MapReduce Programming Model* [11] is an implementation of MapReduce [12], as proposed by Google, for .NET and integrated with Aneka. MapReduce is originated by two functions from the functional language: *map* and *reduce*. The map function processes a key/value pair to generate a set of intermediate key/value pairs, and the reduce function merges all intermediate values associated with the same intermediate key. This model is particular useful for data intensive applications.

The MapReduce Programming Model provides a set of client APIs that allow developers to specify their map and reduce functions, to locate the input data, and whether to collect the results if required. In order to execute a MapReduce application on Aneka, developers need to create a MapReduce application, configure the map and reduce components, and – as happens for any other programming model – submit the execution to Aneka.

MapReduce is good example for the flexibility of the architecture of Aneka in supporting different programming abstractions. With MapReduce the tasks are not created by the user, as with the other supported programming models, but by the MapReduce runtime itself. This peculiarity of the model is hidden within the internal implementation of MapReduce, and it is transparently supported by the infrastructure.

### *3.3. Extending Aneka*

Aneka has been designed to support multiple programming models and its service oriented architecture allows for the integration of additional services. Adding a new programming model then becomes then as easy as integrating a set of services in the Aneka middleware. The support for a specific programming model in Aneka requires the implementation of the following components:

- Abstractions for application composition;
- Client components;
- Scheduling and execution components;

Abstractions define the user view of the programming model, while the other components are internally used by the middleware to support the execution. Aneka provides a default implementation of all these components that can be further specialized to address the specific needs of the programming model. The implementation effort required to integrate a new programming model within Aneka strictly depends on the features of the programming model itself. In order to simplify this process Aneka provides a set of services that can be reused by any model. These are application store, file transfer management, resource reservation, and authentication.

Another way of implementing a new programming model is extending one of the pre-existing models and simply adding the additional features that are required. This could be the case of a workflow implementation on top the Task Model.

### *3.4. Parameter Sweeping Based Applications*

Aneka provides support for directly running existing application on the Cloud without the need of changing their execution logic or behavior. This opportunity can be exploited when the behavior of the application is controlled by a set of parameters representing the application input data. In this case, the most common scenario is characterized by applications that have to be run multiple times with a different set of values for these parameters. Generally, all the possible combinations of parameter values have to be explored. Aneka provides a set of APIs and tools through which it is possible to leverage multiple executions on the Aneka Cloud. These are respectively the Parameter Sweeping APIs and the Design Explorer.

The Parameter Sweeping APIs are built on top of the Task Programming Model and provide support for generating a collection of tasks that will cover all possible combinations of parameter values that are contained in a reference task. The Aneka SDK includes some ready to use task classes that provide the basic operations for composing the task template: execute an application, copy, rename, and delete a file. It also provides an interface that allows developers to create task classes supporting parameter sweeping.

The Design Explorer is a visual environment that helps users to quickly create parameter sweeping applications and run it in few steps. More precisely, the Design Explorer provides a wizard allowing users to:

- Identify the executable required to run the application;
- Define the parameters that control application execution and their domains;
- Provide the required input files for running the application;
- Define all the output files that will be produced by the application and made available to the user;
- Define the sequence of commands that compose the task template that will be run remotely;

Once the template is complete, the Design Explorer allows the user to directly run it on Aneka Clouds by using the parameter sweeping APIs. Different visualizations are provided and statistics collected by the environment in order to monitor the progress of the application.

## 4. Cloud Maintenance and Monitoring

Aneka provides a platform on top of which it is possible to develop applications for the Cloud. The Software Development Kit addresses all the needs from a development point of view but it is just a part of the feature set required by a Cloud Computing platform. Essential in this case is the support for monitoring, managing, maintaining, and setting up computing clouds. These operations are exposed by the management API and the Platform Abstraction Layer on top of which all the management tools and interfaces have been designed. Of a particular interest are the Management Studio and the web management interfaces.

The Management Studio is an important tool for system administrators. It is a comprehensive environment that allows them to manage every aspect of Aneka Clouds from an easy to use graphical user interface. Since Clouds are constituted of hundreds and even thousands of machines both physical and virtual, it is not possible to reach and setup each single machine by hand. Having a tool that allows remote and global management is then a basic requirement. Briefly, the set of operations that can be performed through the Management Studio are the following:

- Quick setup of computing clouds;
- Remote installation and configuration of nodes;
- Remote control of containers;
- System load monitoring and tuning.

Besides the remote control features, which dramatically simplify the management of the Cloud, it is important to notice the support for viewing the aggregate dynamic statistics of Aneka Clouds. This helps administrators to tune the overall performance of the Cloud. It is also possible to probe each single node and collect the single performance statistics: the CPU and memory load information is collected from each container and by inspecting the container configuration it is possible to identify bottlenecks in the Cloud. As the entire framework, the Management Studio has been designed to be extensible: it is possible to add new features and new services by implementing management plugins that are loaded into the environment and get access to the Cloud.

The Management Studio is not the only tool available for controlling Aneka Clouds. The framework also provides a set of web interfaces that provide a programmatic management of Aneka. Currently, only a restricted set of features – resource reservation and negotiation, task submission, and monitoring – is available through web services, while the others are still under development and testing.

## 5. Case Studies

Aneka has been used either in the academic field or in the industry as a middleware for Cloud Computing. In this section we will briefly present some case studies that span from the scientific research to the manufacturing and gaming industry. In all of these cases Aneka has successfully contributed to solve the scalability issues faced and to increase the performance of the applications that leverage the Cloud for their computation needs.

*5.1. Scientific Research*

Aneka has been used to provide support for distributed execution of evolutionary optimizers and learning classifiers. In both of the cases a significant speed up has been obtained compared to the execution on a single local machine. In both of the cases an existing legacy application has been packaged to run in a distributed environment with the aid of a small software component coordinating the distributed execution.

*5.1.1. Distributed Evolutionary Optimization: EMO*

EMO (Evolutionary Multi-objective Optimizer) [13] is an evolutionary optimizer based on genetic algorithms. More precisely, it is a variation of the popular NSGA-II algorithm [14] that uses the information about how individuals are connected to each other – that represents the topology of the population – to drive the evolutionary process. A distributed version of EMO has been implemented on top of Aneka to reduce the execution time of the algorithm and improve the quality of the solutions.

Genetic algorithms [15] are iterative heuristics exploiting the concepts of individual, population, and genes, to define evolving optimizers. These tune their parameters by using mutation, crossover, and mating between individuals, which represent specific points in the solution space. Genetic algorithms have a brute force approach and generally require a large number of iterations to obtain acceptable results. These requirements become even more important in the case of EMO: in order to take advantage of the topology information a large number of individuals and iterations of the algorithms are required. The search for good solutions could require hours, and in the worst case up to one day, even for benchmark problems.

In order to address this issue a distributed implementation of EMO on top of Aneka has been developed [16]. The distributed version of EMO adopts a "*divide and conquer"* strategy and partitions the original population of individuals into smaller populations which are used to run the EMO algorithm in parallel. At the end of each parallel evaluation the results are merged and the next iteration starts. This process is repeated for a predefined number of times.

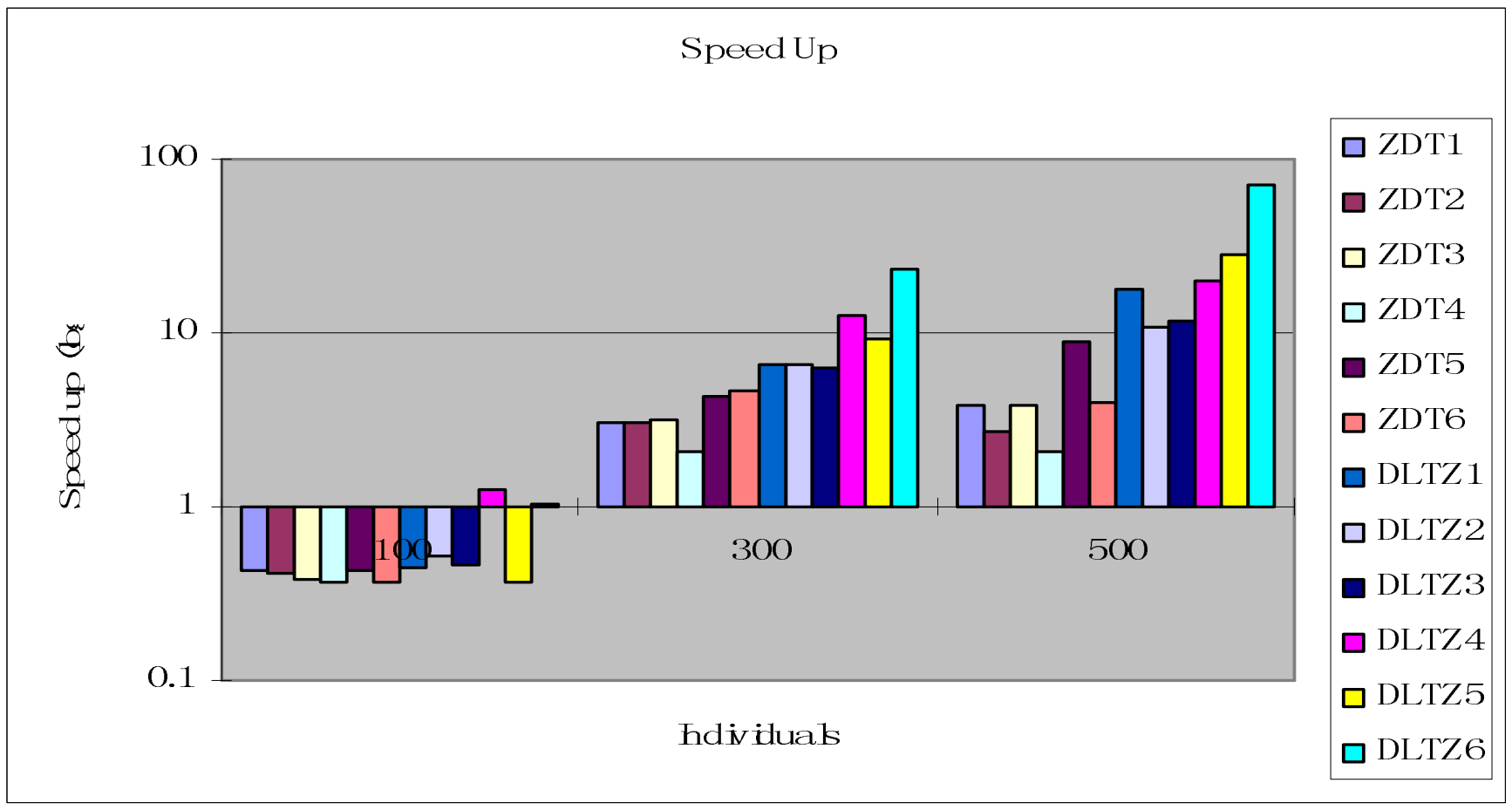

**Figure 6.** Speedup of EMO on Aneka.

Figure 6 and Figure 7 shows the results of running the EMO optimizer on Aneka Clouds for a set of well known benchmark problems ([17] and [18]). The optimization functions used for benchmarking the distributed execution are: ZDT1 to ZDT6, and DLTZ1 to DLTZ6. For each of the optimization functions tested, the graphs respectively show the speedup and the overhead generated while running the optimizer on the Aneka Cloud. It is interesting to notice that for a small number of individual there is no advantage in leveraging the execution on Aneka Clouds. As previously introduced, one of the peculiarities of EMO is the use of topology information for driving the evolutionary process. This information becomes useful when dealing with large number of individuals, at least 1000. As shown by the graphs, the speed up is significant already for 500 individuals, while for 1000 individuals the distribution overhead is completely negligible for all the tested problems.

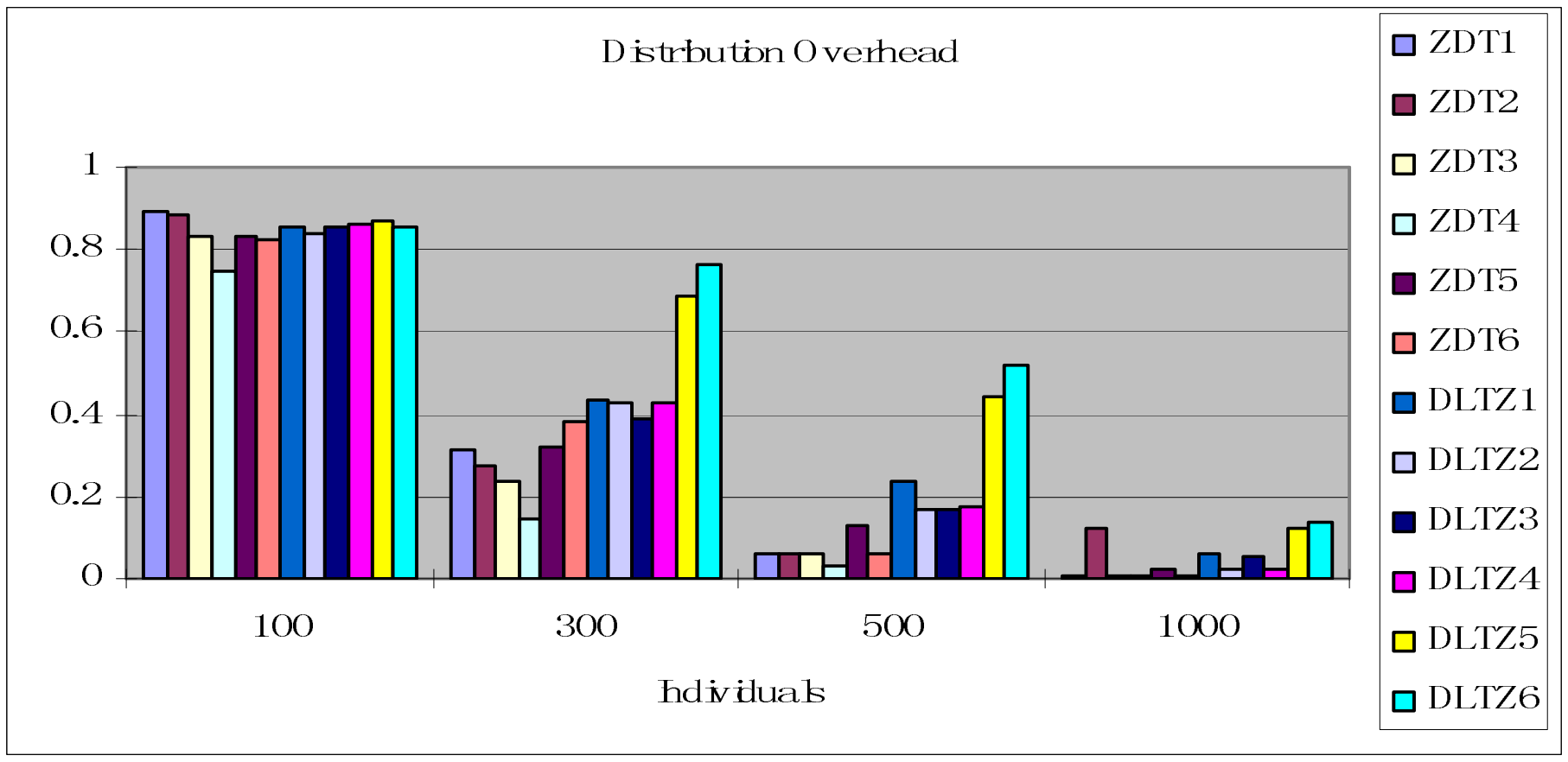

**Figure 7.** Distribution overhead of EMO on Aneka.

### *5.1.2. Distributed Learning Classifiers for Bioinformatics: XCS*

Classifier systems are software systems implementing a function that maps a given attribute set x to a class y. In most of the cases there is no analytic expression for the mapping function. Hence, classifiers use heuristics methods to mimic expected behavior of the mapping function. In particular Learning Classifier Systems (LCS) [19] learn from the input data the structure of the mapping function and adopts genetic algorithms to evolve the set of rules that provides the best performance of the classifier. Several classifiers are derived from LCS. Among these, the eXtended Classifier System (XCS) [20] is popular for the accuracy of the classifiers obtained.

Classifier systems are compute intensive algorithms whose execution time strongly depends on the number of attributes used to classify the samples of a given dataset. Large datasets or simply small datasets with a large number of attributes cause long execution times. In the field of bioinformatics, some specific large datasets containing huge amount of information are used as databases for identifying diseases or finding interesting patterns. Within this context, learning classifiers can be applied to learn from existing classified datasets in order to evolve into classifiers that can support the classification of unseen datasets. The drawback of this approach is that the learning process can last days and does not produce good classifiers. In this scenario the need of having a fast learning process can help bioinformatics researchers to properly tune their classifiers in a reasonable time frame.

In order to reduce the time spent in the learning process of XCS classifiers a distributed implementation based on Aneka has been provided. In particular, a distributed version of XCS has been tuned to support the diagnosis of breast cancer disease by learning from Gene Expression datasets. In order to distribute the learning process the initial dataset has been partitioned into sections that have been used to evolve into different classifiers in parallel for a predefined number of iterations. At the end of each of the iterations the results obtained from each classifier are merged according to different strategies to propagate the good results. The preliminary results have shown that the use of Aneka has contributed to reduce the execution time of the learning process to the twenty percent of the execution on a single machine.

### *5.2. Manufacturing and Gaming Industry*

Besides the research field, Aneka has been used to support real life applications and to address scalability issues in the manufacturing and gaming industry. In particular, the load generated by the rendering of train models and the online processing of multiplayer game logs have been leveraged on a private Aneka Cloud.

#### *5.2.1. Distributed Train Model Rendering: GoFront Group*

GoFront Group is China's premier and largest nationwide research and manufacturing group of rail electric traction equipment. Its products include high speed electric locomotives, metro cars, urban transportation vehicles, and motor train sets. The IT department of the group is responsible for providing the design and prototype of the products including the high speed electric locomotives, metro cars, urban transportation vehicles, and motor trains. The raw designs of the prototypes are required to be rendered to high quality 3D images using the Autodesk rendering software called Maya. By examining the 3D images, engineers are able to identify any potential problems from the original design and make the appropriate changes.

The creation of a design suitable for mass production can take many months or even years. The rendering of three dimensional models is one of the phases that absorb a significant amount of time since the 3D model of the train has to be rendered from different points of views and for many frames. A single frame with one camera angle defined can take up to 2 minutes to render the image. The rendering of a complete set of images from one design require three days. Moreover, this process has to be repeated every time a change is applied to the model. It is then fundamental for GoFront to reduce the rendering times, in order to be competitive and speed up the design process.

In order to face this problem, a private Aneka Cloud has been set up by using the existing desktop computers and servers available in the IT department of GoFront. Figure 8 provides an overall view of the installed system. The setup is constituted by a classic master slave configuration in which the master node concentrates the scheduling and storage facilities and thirty slave nodes are configured with execution services. The task programming model has been used to design the specific solution implemented in GoFront. A specific software tool that distributes the rendering of frames in the Aneka Cloud and composes the final rendering has been implemented to help the engineers at GoFront. By using the software, they can select the parameters required for rendering and perform the computation on the private cloud. Figure 9 illustrates the speed up obtained by distributing the rendering phase on the Aneka Cloud, compared to the previous set up constituted by a single four-core machine. As it can be noticed, by simply using a private cloud infrastructure that harnessed on demand the spare cycles of 30 desktop machines in the department, the rendering process has been reduced from days to few hours.

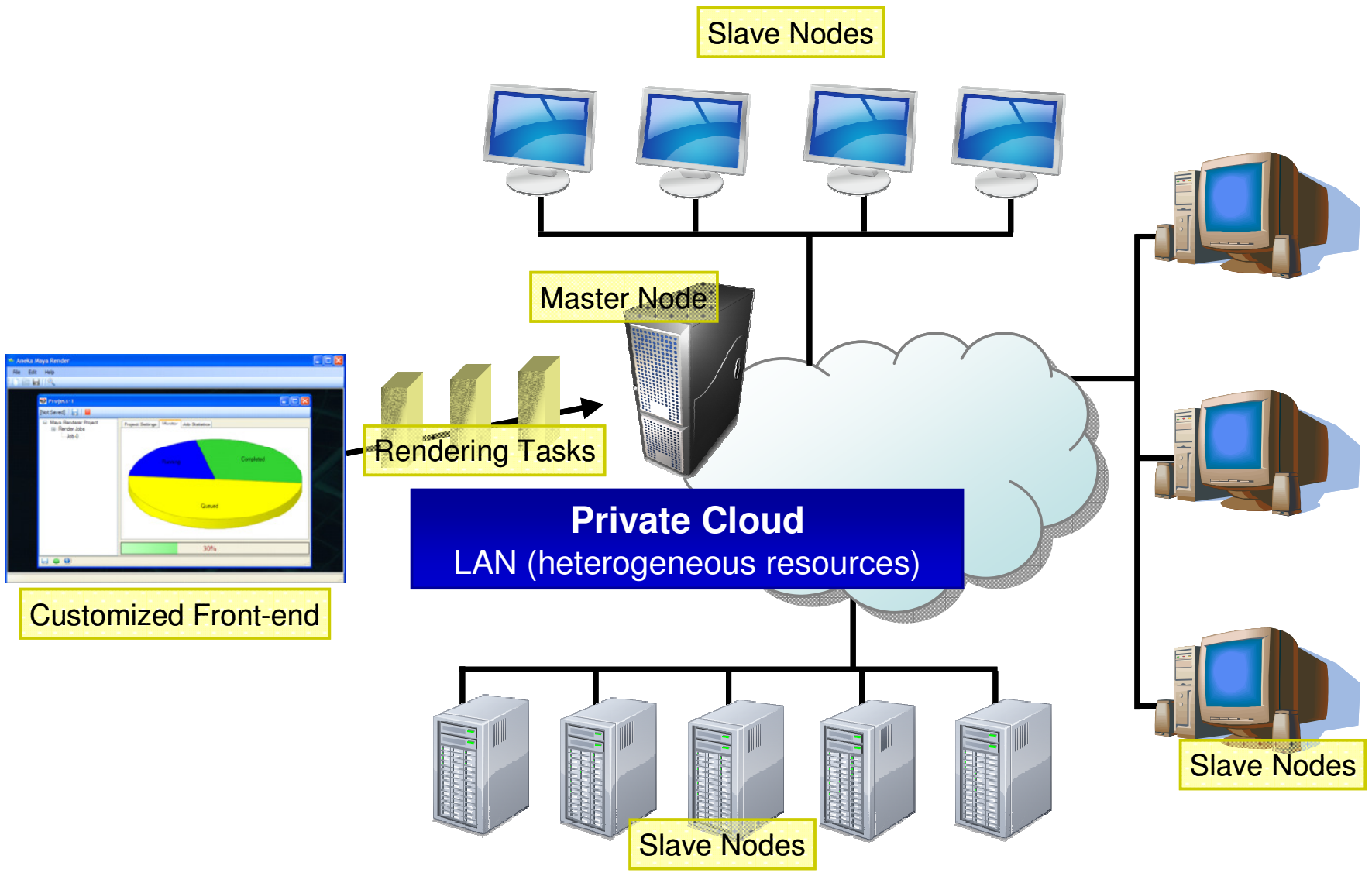

**Figure 8.** Cloud setup at GoFront.

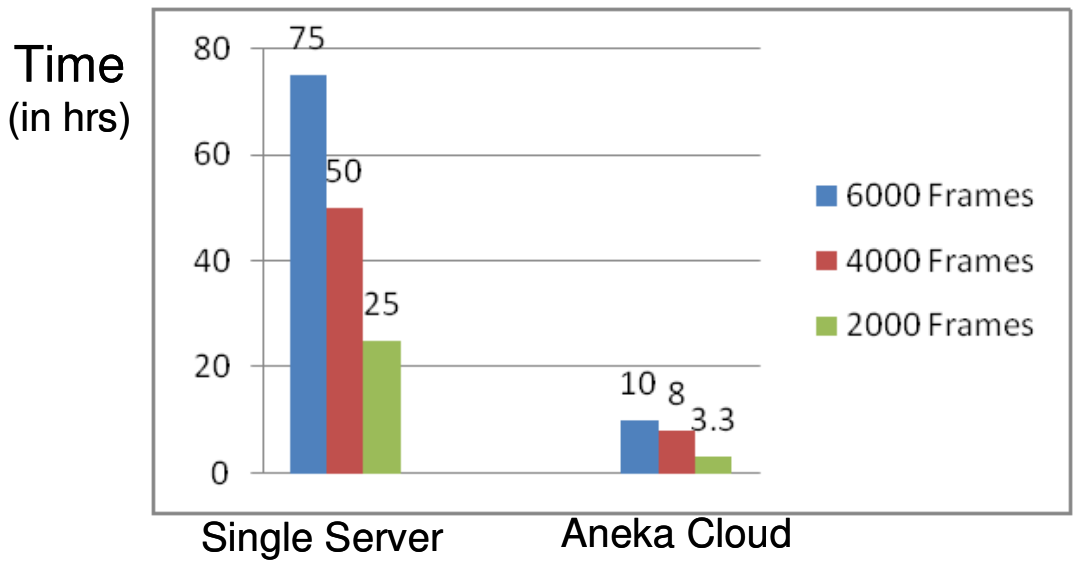

**Figure 9.** Speed up of the rendering process.

### *5.2.2. Distributed Log Processing: TitanStrike Gaming*

TitanStrike Gaming provides an online community for gamers, accessible through a web portal, where they can register their profile, select their preferred multiplayer game, and play on line matches by joining a team. The service provided by TitanStrike is not providing facilities for online gaming, but building a community around them where players can keep and show their statistics and challenge each other. In order to provide such services, the processing of game logs, is fundamental.

An online multiplayer game is generally characterized by a game server that controls one or more matches. While a match is running, players can join and play and the information of everything happening in the game is dumped into the game logs that are used as medium for updating the status of the local view of the game of each player. By analyzing the game logs it is then possible to build the statistics of each player. Game servers generally provide an end point that can be used to obtain the log of a specific game. A single log generates information with a low frequency since the entire process is driven by humans. But in case of a portal for gaming, where multiple games are played at the same time and many players are involved in one match, the overload generated by the processing of game logs can be huge and scalability issues arise.

In order to provide a scalable infrastructure able to support the update of statistics in real time and improve their user experience, a private Aneka Cloud has been set up and integrated into the TitanStrike portal. Figure 10 provides an overall view of the cloud setup. The role of the Aneka Cloud is to provide the horse power required to simultaneously process as many game logs as possible by distributing the log parsing among all the nodes that belong to the cloud. This solution allows TitanStrike to scale on demand when there are flash crowds generated by a huge numbers of games played at the same time.

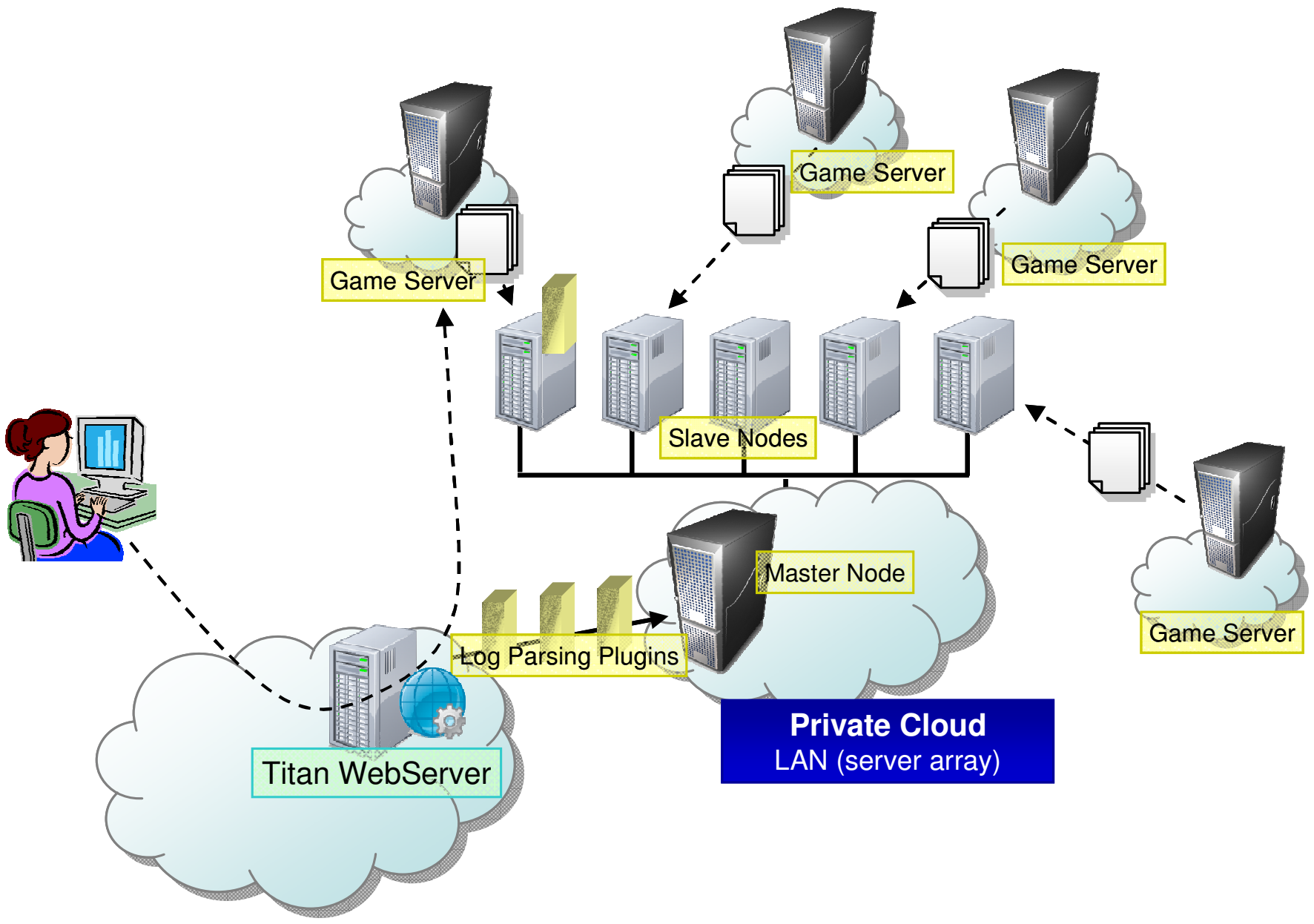

**Figure 10.** Cloud set up at TitanStrike.

## 6. Conclusions and Future Directions

In this book chapter we have presented Aneka, a framework providing a platform for cloud computing applications. As discussed in the introduction there are different solutions for providing support for Cloud Computing. Aneka is an implementation of the Platform as a Service approach, which focuses on providing a set of APIs that can be used to design and implement applications for the Cloud.

The framework is based on an extensible and service oriented architecture that simplifies the deployment of clouds and their maintenance and provides a customizable environment that supports different design patterns for distributed applications. The heart of the framework is represented by the Aneka Container which is the minimum unit of deployment for Aneka Clouds and also the runtime environment for distributed applications. The container hosts a collection of services that perform all the operations required to create an execution environment for applications. They include resource reservation, storage and file management, persistence, scheduling, and execution. Moreover, services constitute the extension point of the container which can be customized to support specific needs or scenarios.

By using services different programming models have been integrated in Aneka. A programming model is a specific way of expressing the execution logic of distributed applications. It provides some familiar abstractions that developers can use to define the execution flow of applications and its component. From an implementation point of view a programming model also includes a collection of services – more precisely scheduling and execution services – that make possible its execution on top of Aneka Clouds. Aneka provides a reference model for implementing new programming models and the current release supports three different programming models: independent bag of tasks, distributed threads, and MapReduce. In order to simplify the development with Aneka a Software Development Kit contains ready to use samples, tutorials, and a full API documentation which helps starting to investigate the large range of features offered by the framework.

Aneka also provides support for deploying and managing clouds. By using the Management Studio it is possible to set up either public or private clouds, monitor their status, update their configuration, and perform the basic management operations. Moreover, a set of web interfaces allows to programmatically managing Aneka Clouds.

The flexibility of Aneka has been demonstrated by using the framework in different scenarios: from scientific research, to educational teaching, and to industry. A set of case studies representing the success stories of Aneka has been reported to demonstrate that Aneka is mature enough to address real life problems used in a commercial environment.

Aneka is under continuous development. The development team is now working on providing full support for the elastic scaling of Aneka Clouds by relying on virtualized resources. Initial tests have been successfully conducted in using Amazon EC2 as a provider of virtual resources for Aneka. This feature, and the ability of interacting with other virtual machine managers, will be included in the next version of the management APIs that will simplify and extend the set of management tasks available for Aneka Clouds.

## Acknowledgements

The authors would like to thank Al Mukaddim Pathan and Dexter Duncan for their precious insights in organizing the contents of this chapter.